\documentclass[aps,prb,twocolumn,superscriptaddress]{revtex4-1}

\usepackage{hyperref} % hyperreferences
\usepackage{graphicx}
\usepackage{amsmath}
\usepackage{amssymb}

\DeclareGraphicsExtensions{.png,.jpg,.eps}
\usepackage{xcolor}

\begin{document}

\author{J. L. Lado}
\affiliation{Department of Applied Physics, Aalto University, Espoo, Finland}
\affiliation{Institute for Theoretical Physics, ETH Zurich, 8093 Zurich, Switzerland}

\author{M. Sigrist}
\affiliation{Institute for Theoretical Physics, ETH Zurich, 8093 Zurich, Switzerland}

\title{
	        Detecting non-unitary multiorbital superconductivity with Dirac
        points at finite energies
}

\begin{abstract}
	        Determining the symmetry of the order parameter of unconventional
        superconductors remains a recurrent topic and non-trivial task in the field of strongly correlated electron
        systems. 
	Here we show that the behavior of Dirac points away from the Fermi
	energy is a potential tool to unveil the orbital structure
        of a superconducting state.
	In particular, we show that gap openings
        in such Dirac crossings are a signature of non-unitary multiorbital
	superconducting order. Consequently, also spectral features at higher
	energy
        can help us to identify broken symmetries of superconducting phases and
	the orbital structure of non-unitary states. Our results
	show how angle-resolved photo-emission spectroscopy measurements can be
	used to detect
	non-unitary
        multiorbital superconductivity.
\end{abstract}

\date{\today}

\maketitle

\section{Introduction}

Superconductors belong to the most intriguing states of matter, 
through their rich phenomenology, the range of underlying physical mechanisms,
their different solid-state realizations 
and ultimately, their potential technological applications. 
Nowadays we label superconductors as conventional or unconventional depending on the symmetry properties of their Cooper pairing states.
Conventional includes all superconductors realizing Cooper pairs of highest
symmetry, whose wave function is essentially structureless and
spin-singlet. In contrast, the term unconventional encompasses all other cases, in particular, those with superconducting order parameters which
violate additional symmetries such as lattice, time-reversal or orbital symmetries, or form topologically non-trivial phases \cite{RevModPhys.47.331,RevModPhys.63.239,Sato2017}.
Strongly correlated electron systems provide an especially fertile ground for
the unconventional superconductivity as they offer a large variety of
low-energy
fluctuations to favoring different Cooper pairing channels \cite{RevModPhys.81.1551,RevModPhys.84.1383}.

\begin{figure}[t!]
\centering
    \includegraphics[width=\columnwidth]{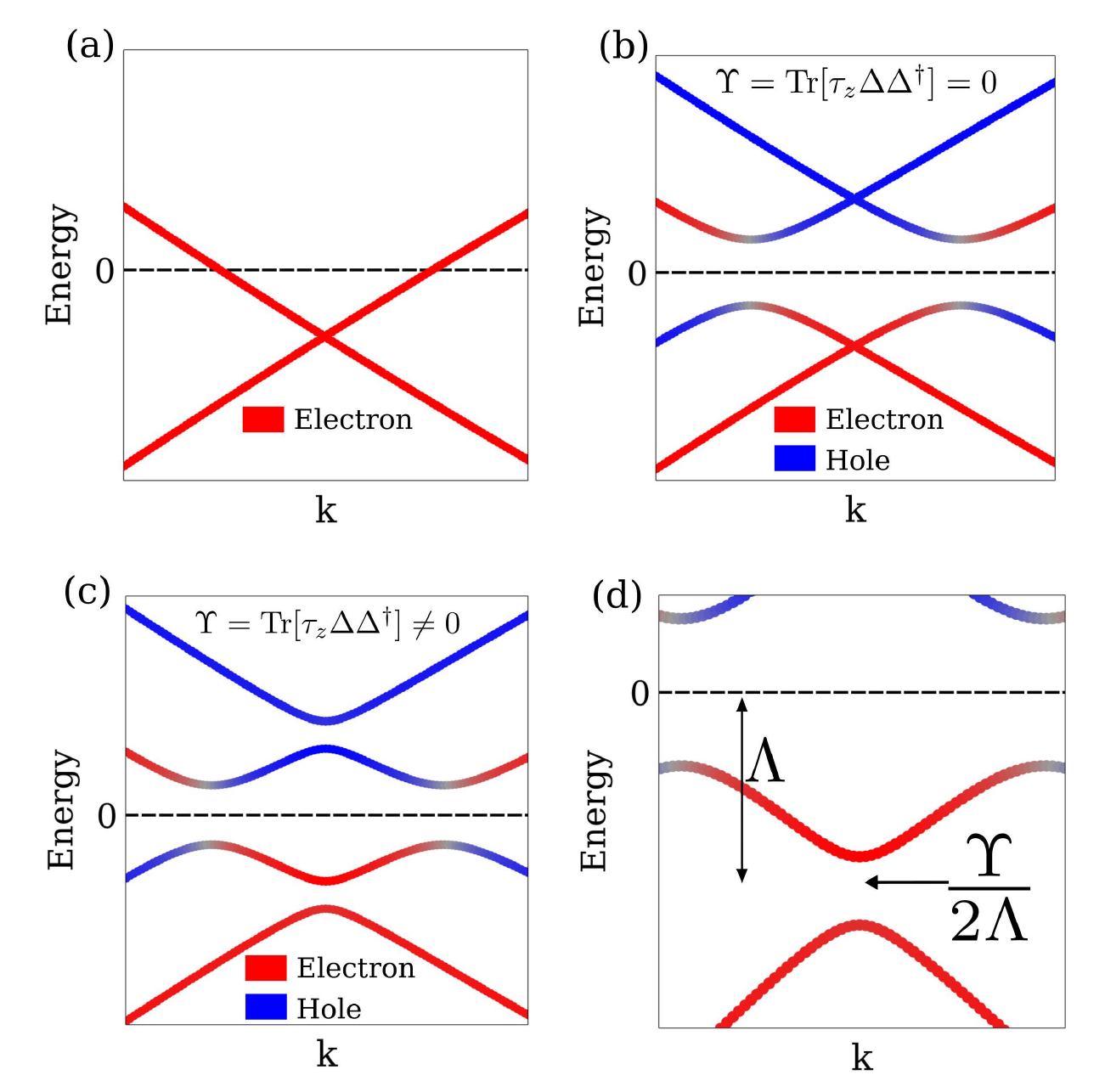}

\caption{
	(a) Sketch of the normal state band structure, featuring
	a Dirac crossing away from the chemical potential.
	(b) Sketch of the band structure in a unitary superconducting state,
	showing that the Dirac point stays closed.
	In stark comparison, the onsite
	of a non-unitary
	superconducting state opens up the Dirac point
	away from the chemical potential as shown
	in panel (c). The magnitude
	of the gap opening at the Dirac crossing
	depends on the non-unitarity $\Upsilon$
	and the distance to the chemical potential
	as shown in panel (d).
}
\label{fig:fig1}
\end{figure}

While the major concepts for unconventional superconductivity are based on
single-band descriptions, multiorbital features increase complexity. For many of
the recently found superconductors a multiorbital approach seems mandatory
to understand the intricate relation of the superconducting phase with the
orbital degrees of freedom and to yield sensible mechanisms for pairing
\cite{RevModPhys.83.1589,Sprau2017}. Very recent experiments on
Fe-chalcogenides\cite{Sprau2017} may show some intriguing aspect of orbital
selectivity of the superconducting phase\cite{Sprau2017}. In angle-resolved
photoemission spectroscopy (ARPES) a conspicuous gap opening of Dirac points
rather far away from the Fermi energy has been observed and interpreted in
terms of spontaneous time-reversal symmetry breaking
stemming from a spontaneous magnetization associated to the
superconducting state
\cite{2019arXiv190711602Z,2019arXiv190601754H}. Motivated by this 
finding, we would like to address
this feature of the remote Dirac points\cite{PhysRevB.92.094517} 
from a general viewpoint and show that
this peculiarity may give 
interesting insights into the multiorbital nature of
the
superconducting phase on a gneral setup. 

Here
we demonstrate that a gap opening in a remote Dirac point
upon entering a superconducting state
indicates the realization of non-unitarity of Cooper pairing states in orbital
space, due to specific forms of orbital selective pairing. The basic
phenomenology is depicted in Fig.\ref{fig:fig1}, showing in
Fig.\ref{fig:fig1} (a) the
normal state spectrum with a Dirac point remote from the Fermi level. After
extending the electronic states by holes to the Nambu space, we obtain the
standard spectrum of a superconductor with a gap around the Fermi energy
(Fig.\ref{fig:fig1}(b,c)). A unitary pairing state leaves the Dirac points
untouched (Fig.\ref{fig:fig1}(b)), while for non-unitary phases these Dirac
points disappear by a gap opening (Fig.\ref{fig:fig1}(c)). In the enlarged
panel Fig. \ref{fig:fig1}(d) we indicate that the size of the gap is
controlled 
by $ \Upsilon $, a measure for 
non-unitarity, as we define below.   
We note that the physical mechanism leading to the superconducting state
can be of various kind, including charge or antiferromagnetic fluctuation,
yet its nature will not affect our analysis.
The paper is organized as follows: in Sec. \ref{sec:dirac} we show
an analysis based on a minimal Dirac equation, in Sec. \ref{sec:tb} we
show how our mechanism applies to a real space model,
in Sec. \ref{sec:interface} we present how the non-unitarity creates
topological interface excitations and in Sec. \ref{sec:con} we
summarize our conclusions.

\section{Non-unitarity in a continuum model}
\label{sec:dirac}

For the theoretical approach to this problem, we first consider a 
generic two-orbital system with a 
two-dimensional Bloch Hamiltonian
hosting a Dirac crossing\cite{Wehling2014}
away from the Fermi level (Fig. \ref{fig:fig1}(a)).
We focus on the
region near the Dirac points of a two-dimensional
band structure using a $ {\mathbf k} \cdot {\mathbf p} $ description
with the spinor basis  $\psi^\dagger_{\mathbf k} = (c^\dagger_{\alpha,\mathbf k},
c^\dagger_{\beta,\mathbf k})$, where
$c^\dagger_{\alpha,\mathbf k}$ and $c^\dagger_{\beta,\mathbf k}$
denote electronic creation operators of Bloch states with momentum $\mathbf k$
and the indices $\alpha,\beta$ label the two orbitals, including also the spin
degrees of freedom.
Then the effective Hamiltonian is written as
$
\mathcal {H}_0^{DP}  (\mathbf k) =
\psi^\dagger_{\mathbf k} H_0^{DP}  (\mathbf k) \psi_{\mathbf k}
$,
where
$H_0^{DP}  (\mathbf k)$ is a $2\times 2$ matrix of the form
\begin{equation}
	H_0^{DP} (\mathbf k) = -\Lambda \tau_0 + \tau_x k_x + \tau_y k_y . 
	\label{heff0}
\end{equation}
The absence of the Pauli matrix $\tau_z$ in Eq. \ref{heff0}
guarantees the existence of a Dirac crossing at the energy $ -\Lambda $. 

We now turn to the superconducting state. For simplicity, we consider here
pairing 
of electrons between time-reversal Kramers partners, which includes 
conventional superconducting states. Note that other channels such as
spin triplet pairing states could be treated in an analogous
way. 
Thus, we extend the Hamiltonian to 
$
\mathcal{H}_{BdG} (\mathbf k)= 
\mathcal{H}_0^{DP} (\mathbf k) + 
\frac{1}{2}
\sum_{i,j \in \alpha,\beta} c^\dagger_{i,\mathbf k} 
c^\dagger_{\bar j,-\mathbf k} \Delta_{ij} (\mathbf k)
+ h.c.
%=
%\frac{1}{2}
%\Psi^\dagger_{\mathbf k} H_{BdG} (\mathbf k) \Psi_{\mathbf k}
$,
%with $\Psi = (
%c^\dagger_{\alpha,\mathbf k},
%c^\dagger_{\beta,\mathbf k},
%c_{\bar \alpha,-\mathbf k},
%c_{\bar \beta,-\mathbf k}
%)$
where $\bar \alpha, \bar \beta$ label the Kramers
time-reversal partners of $\alpha,\beta$ in orbital and spin space.  
We write the $2\times 2$ gap matrix as
\begin{equation}
%\Delta^{\uparrow\downarrow} (\mathbf k) =
\Delta (\mathbf k) = 
	\begin{pmatrix}
\Delta_{\alpha \alpha} (\mathbf k) & \Delta_{\alpha \beta} (\mathbf k)  \\
\Delta_{\beta \alpha} (\mathbf k) & \Delta_{\beta \beta} (\mathbf k) 
\end{pmatrix}
\end{equation}
with 
%\begin{equation}
$%	\begin{matrix}
		\Delta_{\alpha\alpha} (\mathbf k) = 
	\langle c_{\alpha,\mathbf k} 
	c_{\bar \alpha,- \mathbf k}  \rangle
%		\\
	$, $
		\Delta_{\alpha\beta} (\mathbf k) = 
	\langle c_{\alpha,\mathbf k} 
	c_{\bar \beta,- \mathbf k}  \rangle
%		\\
	$, $
		\Delta_{\beta\alpha} (\mathbf k) = 
	\langle c_{\beta,\mathbf k} 
	c_{\bar \alpha,- \mathbf k}  \rangle
%		\\
$, $
	\Delta_{\beta\beta} (\mathbf k) = 
	\langle c_{\beta,\mathbf k} 
	c_{\bar \beta,- \mathbf k}  \rangle
%	\end{matrix}
$
	%\end{equation}
For systems where $s_z$ is a good quantum number, the indices are
$\alpha \equiv (\alpha_0,\uparrow)$
and
$\bar \alpha \equiv (\alpha_0,\downarrow)$ 
with $\alpha_0$ the orbital label, such that 
$\Delta (\mathbf k) \equiv \Delta^{\uparrow\downarrow} (\mathbf k)$
describes opposite spin pairing. For concreteness, we 
continue now with this orbital spin labeling, although 
more general Kramers pairs yield the same behavior. 
With the Nambu spinor
$\Psi^\dagger_{\mathbf k} =
(
c^\dagger_{\alpha_0,\uparrow,\mathbf k},
c^\dagger_{\beta_0,\uparrow,\mathbf k},
c_{\alpha_0,\downarrow,-\mathbf k},
c_{\beta_0,\downarrow,-\mathbf k}
)$, the
Bogoliubov-de-Gennes (BdG)
Hamiltonian $H_{BdG}$
takes the form
$\mathcal{H}_{BdG} = 
\frac{1}{2}
\Psi^\dagger_{\mathbf k} H_{BdG}  (\mathbf k) \Psi_{\mathbf k} + h.c.$, where
$H_{BdG}  (\mathbf k)$ is the following 
$4\times4$ matrix

\begin{equation}
	H_{BdG}  (\mathbf k) = \begin{pmatrix}
H^{DP}_0 (\mathbf k)  & \Delta (\mathbf k)\\
		\Delta (\mathbf k) & - H^{DP}_0 (-\mathbf k)
\end{pmatrix}
\end{equation}
Using this Hamiltonian, 
we now derive the corrections induced by the Cooper pairing
at the Dirac point of Eq. \ref{heff0}.
A Schrieffer-Wolff transformation
equivalent to second order perturbation theory leads to the
effective Hamiltonian around the Dirac point $\mathbf k = \mathbf K$ of the form

\begin{equation}
	H^{DP} = H^{DP}_0 +
\frac{
 \Delta \Delta^\dagger 
}
{2\Lambda}
	\label{eq:deff}
\end{equation}
with $\Delta \equiv \Delta (\mathbf K)$,
valid in the limit $|\Lambda| \gg \text{max}(|\Delta|$).

\begin{figure}[t!]
\centering
    \includegraphics[width=\columnwidth]{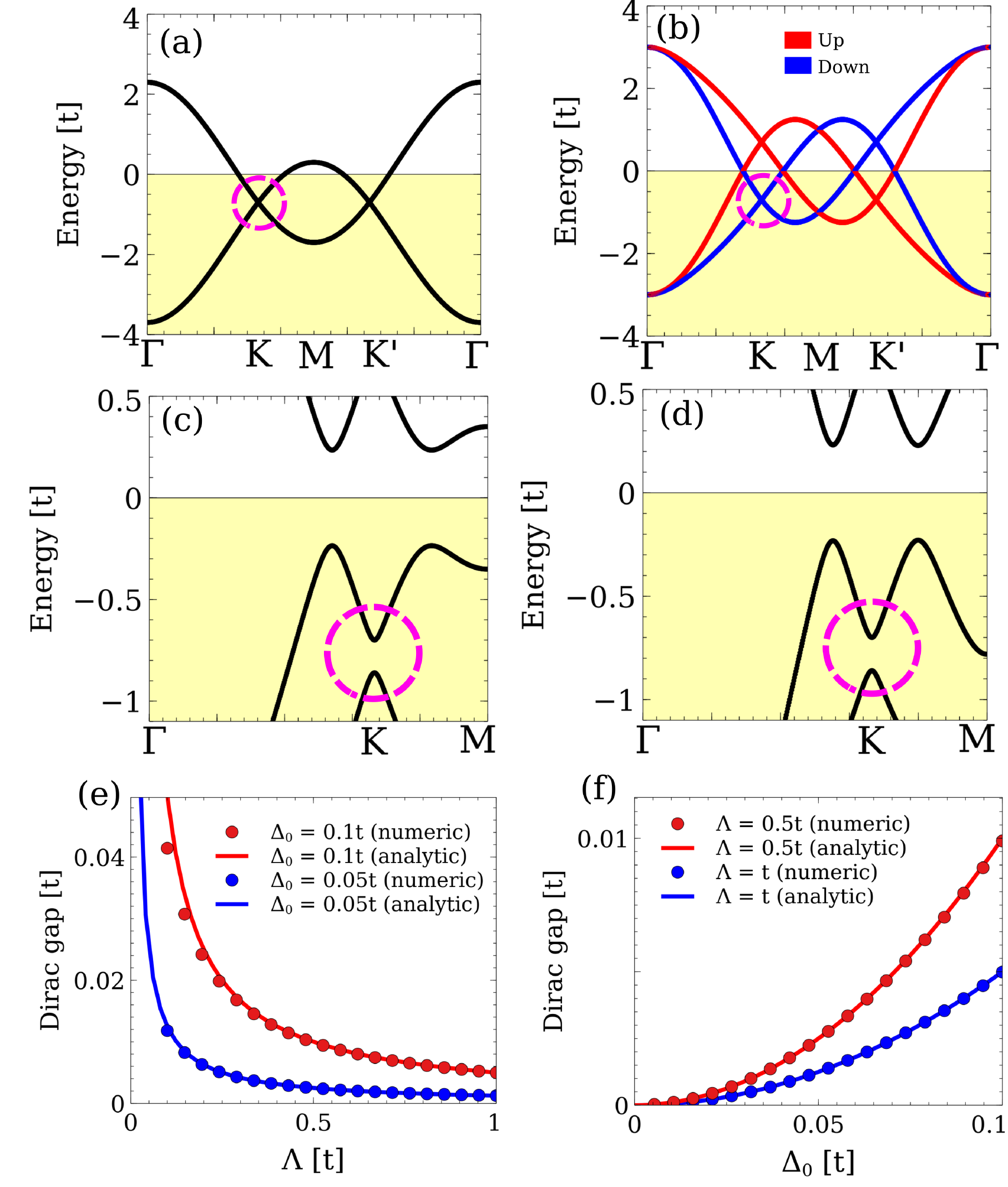}

\caption{
	Normal state band structure (a,b) for a 
	Dirac band system in the normal state,
	showing Dirac crossings away from the chemical potential.
	Panel (a) shows a doped honeycomb
	lattice without spin orbit-coupling,
	whereas panel (b) shows
	a half filled honeycomb system with non-centrosymmetric
	spin-orbit coupling (b).
	Panel (c) shows the BdG spectra
	in the presence of non-unitary superconducting term 
	for (a), and panel (d) the analogous spectra for the case (b).
	In both scenarios, it is observed that the non-unitary
	superconducting state drives a gap opening at the Dirac point.
	Panels (e,f) show a comparison between
	the numeric and analytic results for Dirac gap opening.
	In particular, panel (e) shows the Dirac gap
	as a function of the distance to the chemical potential
	$\Lambda$ (e), and panel (f) the Dirac gap
	as a function of the non-unitary superconducting term.
	We took $\Lambda_1=0.7t$ and $\Lambda_2=0$ for (a,c),
	$\Lambda_1=0.0$ and $\Lambda_2=0.7t$ for (b,d),
	$\Lambda_2=0$ for panels (e,f)
	and $\Delta_0 = 0.71\Lambda $ for (c,d).
}
\label{fig:fig2}
\end{figure}

As a $2\times 2$ matrix,
the correction to the Dirac Hamiltonian in Eq. \ref{eq:deff} 
can be decomposed into the identity $\mathcal{I}$ 
and Pauli matrices $\tau_i$ as
$
\frac{\Delta \Delta^\dagger}{2\Lambda} =
\gamma_0 \mathcal{I} + 
\gamma_x \tau_x +
\gamma_y \tau_y + 
\gamma_z \tau_z
$
with $\gamma_0,\gamma_x,\gamma_y,\gamma_z$ real numbers, where
$ \gamma_x = \gamma_y = \gamma_z = 0 $ defines a unitary pairing state. 
Generally, this yields an effective Hamiltonian around the original
Dirac point that can be again written in terms the identity $\mathcal{I}$
and Pauli matrices $\tau_i$ as,

\begin{equation}
	H^{DP} =
	-\bar \Lambda \tau_0 +
	\bar p_x \tau_x +
	\bar p_y \tau_y +
	\gamma_z \tau_z 
	\label{eq:heff}
\end{equation}

with 
$\bar \Lambda = \Lambda - \gamma_0$,
$\bar p_x = p_x + \gamma_x$ and
$\bar p_y = p_y + \gamma_y$.
This highlights that a non-unitary superconducting term
modifies the Dirac crossing away from the chemical potential. 
As a result, the existence of non-unitary pairing can
be generically observed by analyzing the behavior of Dirac point
in the superconducting state. 
For non-vanishing $\gamma_x \ne 0$ or $\gamma_y \ne 0$, the onset of superconductivity 
will shift the Dirac point in reciprocal space.
But more importantly, for $\gamma_z \ne 0$ in Eq. \ref{eq:heff} a gap 
opens at the Dirac point. Generally, the condition to open a gap at the Dirac point can be written in terms
of the non-unitarity parameter $\Upsilon$
of the gap matrix as
$\Upsilon =
	\sum_{\nu=\alpha,\beta} 
	[\Delta_{\alpha,\nu} \Delta^*_{\alpha,\nu} -
	\Delta_{\beta,\nu} \Delta^*_{\beta,\nu} ]
$,
which in matrix form reads
\begin{equation}
\Upsilon = \text{Tr}[\tau_z \Delta \Delta^\dagger]\ne 0
\label{upsilon}
\end{equation}
yielding $\gamma_z = \Upsilon / 4\Lambda $ in Eq. \ref{eq:heff}.
We point out that a similar analysis could be carried out both for 
one-dimensional and three-dimensional systems described by a 
$2\times 2$ Dirac equation. 
However, in the three dimensional case, a gap can not
be induced in the $2\times 2$ Dirac Hamiltonian, as any correction
will only create a momentum shift.

We examine now a few examples based on this criterion. 
We start with the unitary states. First the intra-orbital state
given by $\Delta (\mathbf k) = \Delta_0 \tau_0 f(\mathbf k)$
which for $ f (\mathbf k) = const. $ corresponds to a conventional
superconductor. This leads obviously to $ \Upsilon = 0 $. The same
is true of an inter-orbital pairing state like $\Delta (\mathbf k) = \Delta_0 \tau_x f(\mathbf k)$. 
A linear combination of these two states is non-unitary with $ \Upsilon = 0 $,
and leads to a shift of the Dirac points.

A straightforward case opening a gap is the
intra-orbital pairing state 
$\Delta (\mathbf k) = 
\Delta_0 f(\mathbf k)
	\begin{pmatrix}
1 & 0  \\
0 & 0
\end{pmatrix}
= 
\Delta_0 f (\mathbf k) (\tau_0+\tau_z)/2
$, which corresponds to pairing restricted to one orbital only
and leads to $ \Upsilon \ne 0 $. Such a state breaks
orbital symmetry, and it is expected to appear close to a quantum
critical point to an orbital ordered instability. 
An exemplary non-unitary state for inter-orbital pairing 
is 
$\Delta = \Delta_0 f(\mathbf k)
	\begin{pmatrix}
0 & 1  \\
0 & 0
\end{pmatrix}
= 
\Delta_0 f(\mathbf k)
(\tau_x + i \tau_y)/2
$, which shows spontaneous 
time-reversal symmetry breaking in orbital space
and realizes an antiferromagnetic superconducting state
breaking spin rotational symmetry.
This state is expected to appear close to a quantum
phase transition to an antiferromagnetic
instability.
We finally emphasize that in the previous description we have
not assumed any momentum space structure of the superconducting state,
and as a result this argument can be applied to arbitrary momentum
superconducting structures.

\begin{figure}[t!]
\centering
    \includegraphics[width=\columnwidth]{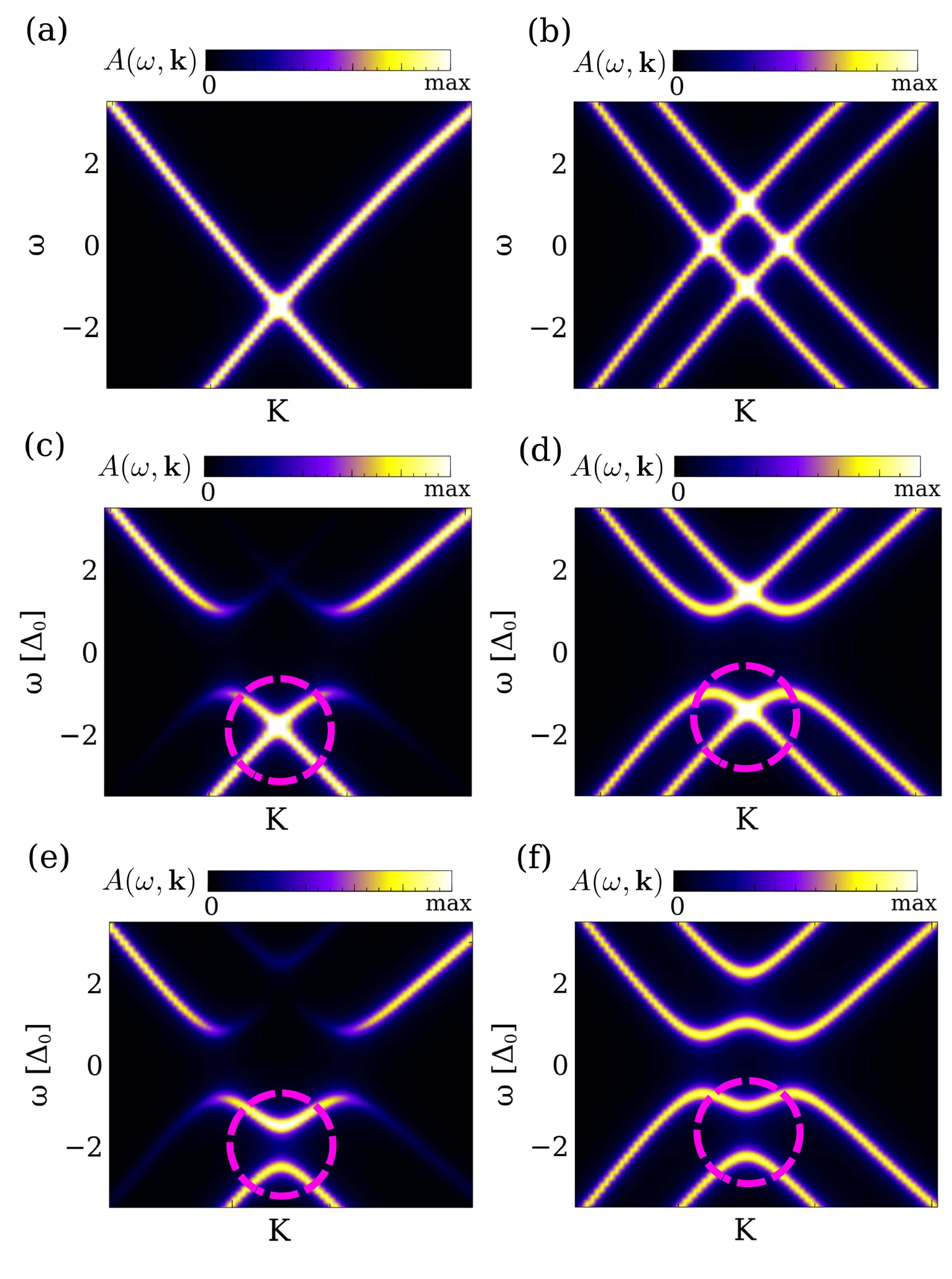}

	\caption{ 
	Electron spectral function $A(\omega,\mathbf k)$
	for a system
	with a Dirac point
	away from the chemical potential
	due to doping (a,c,e)
	and due to spin-orbit coupling
	(b,d,f).
	Panels (a,b) show the spectral function
	in the normal state,
	depicting a Dirac crossing away form the chemical potential.
	In the presence of unitary orbital superconductivity,
	a gap opens up at the chemical potential,
	but not at the Dirac crossings
	(c,d).
	In contrast, in the presence of a non-unitary
	superconducting state a gap opens up at the
	Dirac crossing (e,f).
	We took $\Lambda_1=1.5 \Delta_0$ and $\Lambda_2=0$
	in (a,c,e), $\Lambda_1=0$ and $\Lambda_2= \Delta_0$ in (b,d,f).
}
\label{fig:fig3}
\end{figure}

\section{Non-unitarity in a real space model}
\label{sec:tb}

We now illustrate this analysis on a specific tight-binding model for the
honeycomb lattice. 
The role of orbitals is now played by the two sublattices. 
The Hamiltonian has the form
\begin{equation}
	\begin{split}
		\mathcal{H}_{TB} = 
	t \sum_{\langle ij \rangle,s} c^\dagger_{i,s} c_{j,s}
	+
	\Lambda_1 \sum_{i,s} c^\dagger_{i,s} c_{i,s} \\ \\
	+ 
\Lambda_2 \frac{i}{3\sqrt{3}}
\sum_{\langle \langle ij \rangle \rangle,s,s'}
\nu_{ij}  \sigma_z^{ss'} \tau_z^{ij} 
c^\dagger_{i,s}
		c_{j,s'} \\ +
	\sum_{i} (
	\Delta^{\uparrow\downarrow}_{i} 
	c_{i,\uparrow} c_{i,\downarrow} + h.c. )
	\end{split}
	\label{eq:tb}
\end{equation}
with $c^\dagger_{is},c_{is}$
are the creation/annihilation operators of electrons on the site $ i $ with spin $s$,
$\langle \rangle$ denotes the summation over nearest neighbors,
$\langle \langle \rangle \rangle$ over next-nearest neighbors, $ \sigma^z $ is
the corresponding spin Pauli matrix and $\nu_{ij}=\pm 1$
for clockwise/anticlockwise hopping within each hexagonal plaquette, whereby an
additional sign difference enters through the orbital Pauli matrix $
\tau_z $,
giving the opposite sign for the two sublattices. 
Note that the $ \Lambda_2 $ term describes
the non-centrosymmetric spin-orbit coupling for next-nearest neighbor hopping, which induces
a momentum-dependent spin-splitting in the band structure with shifted Dirac cones \cite{PhysRevLett.61.2015,PhysRevLett.95.226801}.  The $\Lambda_1$ term
introduces merely the chemical potential which is finite and negative for the electron-doped system. 
Finally, $\Delta^{\uparrow\downarrow}_{i}$ controls the amplitude of the onsite
pairing state, yielding an intra-orbital pairing. 
We will use here the simplest spin singlet s-wave superconducting pairing
to highlight the impact of a non-trivial orbital structure even for a
topologically trivial superconducting state.
Other superconducting momentum structures could be studied analogously.\cite{PhysRevX.9.031025,PhysRevB.98.224509}

The Hamiltonian Eq. \ref{eq:tb} has two simple limiting cases. 
First, for $\Lambda_1 \ne 0$ and $\Lambda_2 = 0$,
the band structure shows two Dirac points below the Fermi level, 
realizing a doped honeycomb lattice (Fig. \ref{fig:fig2}(a)).
Second,  with $\Lambda_1 = 0$ and $\Lambda_2 \ne 0$ the system is
half-filled,
but the Dirac cones are spin split due to the non-centrosymmetric
spin-orbit coupling, and there are again Dirac points away from the Fermi energy
(Fig. \ref{fig:fig2}(b)). 
In the generic case, $\Lambda_1 \ne 0$ and $\Lambda_2 \ne 0$,
Dirac crossings will appear at energies $\Lambda = \Lambda_1 \pm \Lambda_2$.

Now we introduce the non-unitary pairing state $\Delta = \Delta_0(\tau_0 + \tau_z)/2$,
which corresponds to onsite pairing on one of the two sublattices only. For
both limiting cases,
the Dirac point remote from the Fermi energy develops
a gap as seen in Fig. \ref{fig:fig2}(c,d). 
Moreover, we compute the magnitude of this gap
as function of $ \Lambda = \Lambda_1 $ and
$ \Delta_0 $ and compare this with our result for $ \Upsilon / 2 \Lambda $ of Eq.(\ref{upsilon})
demonstrating very good agreement as long as the condition $ | \Lambda | \gg | \Delta_0 | $ is satisfied (Fig. \ref{fig:fig2}(e,f)).  
Note that this pairing state is a so-called pair density wave state and corresponds to a
spontaneous sublattice symmetry breaking. It would be accompanied by a charge density
difference on the two sublattices and is stabilized in systems where superconductivity appears 
close to quantum phase transition to a charge density wave instability.

We now discuss the electron spectra function which is experimentally accessible,
for instance, by ARPES. This can be computed through, 
\begin{equation}
	A(\omega, \mathbf k) = 
	-\frac{1}{\pi} \text{Im} [ \text{Tr} 
	[P_e (\omega - H(\mathbf k) + i0^+)^{-1} ]
\end{equation}
with $H(\mathbf k)$ as the corresponding BdG Hamiltonian in the Nambu representation and $P_e$
the projection operator on the electron sector. The quantity $A(\omega, \mathbf k)$
would be the observable in an ARPES experiment in the superconducting state,
as shown in Fig. \ref{fig:fig3}. We consider again the two limiting cases for
the Dirac cones shifted below the
Fermi energy in Fig. \ref{fig:fig3}(a,c,e) for electron doping 
and in Fig. \ref{fig:fig3}(b,d,f) for spin-orbit coupling. 
The spectrum for the normal state spectrum is shown in Fig. \ref{fig:fig3}(a,b), and for superconducting states
in Fig. \ref{fig:fig3} (c,d,e,f), both for a unitary (Fig. \ref{fig:fig3}(c,d))
and a non-unitary (Fig. \ref{fig:fig3}(e,f)) superconducting state.  The
unitary state is given by
$\Delta^U = \Delta_0 \tau_0$ and the non-unitary one 
by  $\Delta = \Delta_0(\tau_0 + \tau_z)/2$. As anticipated from our discussion above,
again a gap opens at the Dirac points below from the Fermi level only if the pairing
state is non-unitary. In all cases the superconducting gap at the Fermi energy
is 
fully open. 

The real space analysis
demonstrates that the mechanism we presented is robust against corrections to
the Dirac dispersion such as trigonal warping or electron-hole asymmetry,
present in the lattice model but not in the continuum model. As a result, the
lattice model demonstrates that our phenomenology is a robust mechanism that
will apply even in cases beyond a minimal Dirac equation, highlighting that it
will be applicable to dispersions of real compounds. This will be especially
important
in the next section, when we show the emergence of non-trivial
interface excitation on a real-space boundary.

\begin{figure}[t!]
\centering
    \includegraphics[width=\columnwidth]{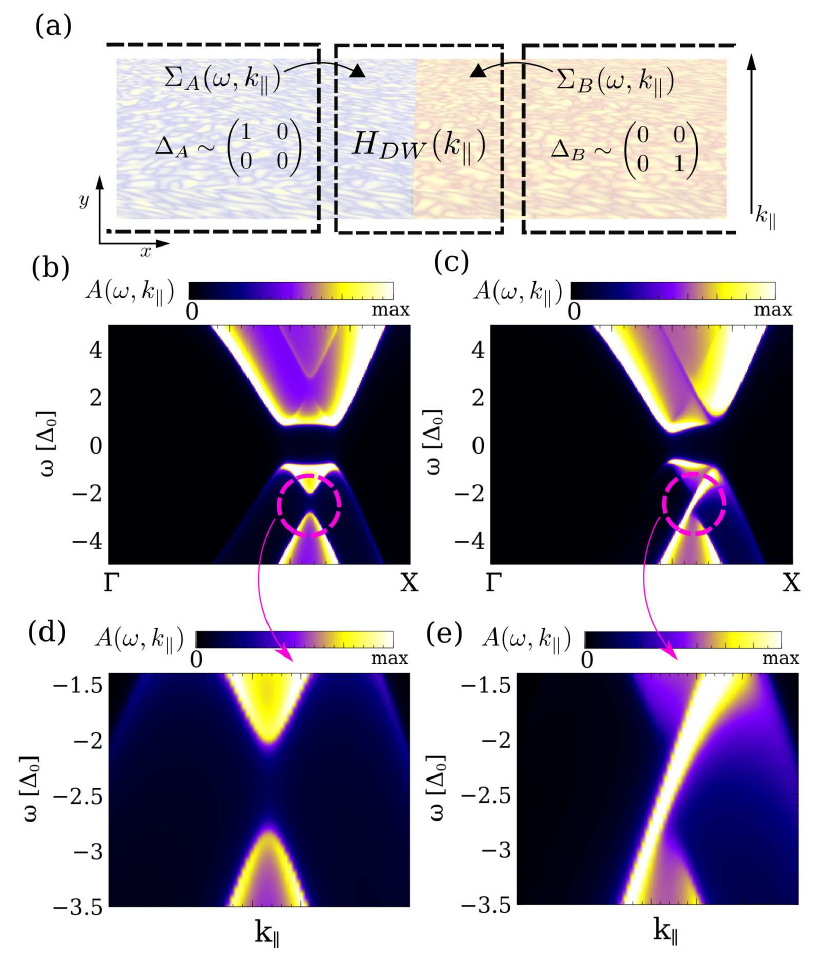}

\caption{
	(a) Sketch of an interface between two superconducting domains with
	different non-unitary orders
	$\Delta_A$ and $\Delta_B$. Panels (b,c) show the bulk spectral
	function of the superconductor in the absence
	of a superconducting domain wall,
	showing the gap opening in the buried Dirac point (b,d).
	In contrast, at the interface between the two
	superconducting domains,
	a chiral excitation appears inside
	the Dirac opening as shown in (c,e).
	We took $\Lambda = 1.5 \Delta_0$.
}
\label{fig:fig4}
\end{figure}

\section{Interface modes associated to non-unitarity}
\label{sec:interface}

Gapped Dirac points are a known source of topological excitations.
In the following, we show how these gaps give rise to in-gap excitations 
at domain walls between superconducting domains, which otherwise
lack any spectroscopic signature at the Fermi energy. For this purpose,
we now take an interface between two degenerate non-unitary
superconducting phases $A$ and $B$, namely
$
\Delta_A = \Delta_0 (\tau_0 + \tau_z)/2
$
and 
$
\Delta_B = \Delta_0 (\tau_0 - \tau_z)/2
$.
We now consider a system with translational
invariance along the domain wall,
direction $\parallel$, with $\Delta_A$ on the left and 
$\Delta_B$ on the right (see Fig. \ref{fig:fig4}(a)).
Here the spectral function at the superconducting 
domain wall can be obtained by 
the Green's function
$
	G(\omega,k_\parallel) = [\omega - H_{DW}(k_\parallel) -
\Sigma_A(\omega,k_\parallel)- \Sigma_B(\omega,k_\parallel)]^{-1}
$
where $k_\parallel$ is the momentum parallel to the domain wall,
and $H_{DW}$ denotes the local Hamiltonian at the interface,
where, for simplicity, we take a sharp transition between the
two superconducting gaps. 
The self-energies $\Sigma_{A,B}(\omega,k_\parallel)$ are  
induced by the semi-infinite superconducting regions, which can
be exactly computed thorugh a frecuency dependent
renormalization group algorithm.\cite{Sancho1985} Then the spectral function
at the interface can be computed by
$A(\omega, k_\parallel) = 
-\frac{1}{\pi}\text{Im} [\text{Tr}[P_e G(\omega,k_\parallel) ] ]
$, which gives access to the local excitations. Physically, the
system we are computing corresponds to an interface between
two semi-infinite superconducting orders with a sharp transition
between, yet this procedure can be
trivially extended to a smooth transition.

Away from the domain wall, the Dirac point shows a gap in the
$\mathbf k$-resolved spectral function (Fig. \ref{fig:fig4}(b,d)). 
At the domain wall, a chiral state appears inside the gap, corresponding to
a chiral Andreev bound state (Fig. \ref{fig:fig4}(c,e)).
The origin of such in-gap excitations
can be easily rationalized by defining the
Chern number associated to the effective Dirac Hamiltonian
Eq. \ref{eq:deff}, 

\begin{equation}
	C = \frac{1}{2} 
	\text{sign} (\Upsilon)
	\text{sign} (\Lambda)
	\label{eq:chern}
\end{equation}
Taking 
$\Delta_0>0$
for $\Delta_A$ and $\Delta_B$,
we obtain 
$C_A=+1/2$ 
and 
$C_B=-1/2$,
which from the index theorem
implies that a single interface state
will appear inside the Dirac gap (Fig. \ref{fig:fig4}(e)) \cite{PhysRevD.13.3398}.
We emphasize that the non-integer nature of the Chern number stems
from the fact that the Dirac equation is defined in a non-compact manifold,
which is associated with taking a Dirac equation
as a low energy approximation.
Interestingly, such interface states in the Dirac
gap do not create an Andreev bound states within
the actual superconducting gap at the Fermi level. 
This shows that such domain walls are invisible
in the actual low-energy spectroscopic gap of the superconductor,
but can be observed by targeting Dirac point openings
away from the chemical potential.
Note that there are no states inside
the Dirac opening in a boundary
with vacuum,\footnote{A zigzag interface with vacuum will show a flat band
of zigzag boundary modes, yet without crossing the gap} such that 
only the gap opening at the Dirac points would be visible in ARPES at a surface.
We finally mention that the real-space model
allows to clearly demonstrate that emergence of interface excitations between
the two superconducting domains by exactly computing such a real space
interface.  This robustly shows that potential corrections to the low energy
model do not spoil the topological interface modes, demonstrating that an
approximate analysis in terms of a low energy Dirac equation can yield
qualitatively faithful results.

\section{Conclusions}
\label{sec:con}

To summarize, we have shown that a gap opening at a Dirac point
away from the Fermi level is a hallmark of a multiorbital non-unitary state.
In particular, our results show that information on the orbital symmetry of the superconducting state
could be directly extracted from an ARPES experiment, providing a simple
spectroscopic technique to detect unconventional non-unitary superconductivity of this kind. 
On the one hand, our methodology can be readily applied to 
iron chalcogenides, where
signatures of orbital-selective states have been observed,\cite{Sprau2017}
and that are known to host remote
Dirac crossings in the band structure \cite{PhysRevB.93.104513}.
On the other hand, our results may potentially provide an alternative
route to characterize the superconducting state of twisted bilayer graphene,
\cite{Cao2018,Yankowitz2019,2019arXiv190306513L}
as it hosts Dirac points 10 meV below
the Fermi level.\cite{PhysRevB.82.121407,PhysRevLett.99.256802,Bistritzer2011,PhysRevLett.123.046601}
Ultimately, our results highlight that ARPES measurements can help to identify
the origin of superconductivity in Dirac compounds. Moreover, it may also
reveal the presence of an
incipient order, such as a charge density wave in our example above, 
which might be involved in the pairing mechanism due to strong fluctuations. 

\

\textbf{Acknowledgements:}
We are grateful to P. Johnson for many helpful discussions on his
experimental results. 
J.L.L. is grateful for financial support from the ETH Fellowship program and
M.S.
from the Swiss National Science Foundation 
through Division II (Grant No. 184739).

\bibliographystyle{apsrev4-1}
\bibliography{biblio}{}

\end{document}